\documentclass[runningheads]{llncs}
\usepackage[T1]{fontenc}
\usepackage[utf8]{inputenc}
\usepackage{babel}
\usepackage{graphicx}
\usepackage{color}
\usepackage{multirow}

\begin{document}

\title{Navigating ICT In-House Procurement in Finland: Evaluating Legal Frameworks and Practical Challenges}
\titlerunning{Navigating ICT In-House Procurement in Finland}
%
\author{Reetta Ghezzi \and
Minnamaria Korhonen \and \\
Hannu Vilpponen \and
Tommi Mikkonen}

\authorrunning{R. Ghezzi et al.}
%
\institute{University of Jyväskylä, Finland\\
\smallskip
\email{reetta.k.ghezzi@jyu.fi}, 
\email{minnamaria.korhonen@gmail.com}, 
\email{hannu.v.vilpponen@jyu.fi},
\email{tommi.j.mikkonen@jyu.fi}
}

\maketitle  
\begin{abstract}
In-house procurement is a controversial issue in the field of public procurement. Simply put, such procurement allows overlooking certain aspects of fair and equal treatment of vendors. This paper presents  qualitative research on in-house ICT procurement within Finnish municipalities. 
Semi-structured interviews were conducted to gather insights from municipal stakeholders. Using grounded theory approach, data analysis shows intricate dynamics between Finnish municipalities and in-house entities associated with them. 
Still, it is clear that the legal framework governing in-house procurement remains intricate and debated.

\end{abstract}
\section{Introduction}
Public sector is a large consumer of ICT systems and services \cite{arrowsmith2002public}.  For example, the Finnish government alone made ICT purchases worth over EUR 1000 million in 2022 \cite{tutkihallintoa_2022}. In addition, Finnish municipalities, joint municipal authorities, and parishes made ICT purchases worth almost EUR 1500 million \cite{tutkihallintoa_2022}. 
With this in mind, the
Public Procurement Directive \cite{directive-procurement} encourages EU Member States to adopt transparent and pro-competitive procurement practices. Public bodies may adopt vast procurement opportunities to achieve these requirements.  
The first option is to publicly tender the purchase \cite{hankintakasikirja_2017}. The second option involves in-house procurement or procurement from other units of the stakeholder, which falls outside the scope of public procurement law, in this case \cite{hankintakasikirja_2017}.

So called in-house companies are owned by public organizations. Their role in public sector procurement has recently attracted a lot of attention, as 
transparency and openness in in-house procurement can be difficult to implement \cite{ghezzi2022role}. Moreover, in-house procurement can also lead to difficulties in obtaining information and data from in-house companies. Finally, legal interpretations of in-house status can be unclear \cite{ghezzi2022role}.

In this paper, we study how much Finnish municipalities rely on in-house procurement and why municipalities do or do not use in-house procurement. Sixteen semi-structured interviews with procurement and ICT key persons in municipalities were used to collect the research data. The interviews were conducted face-to-face or by video conference, whichever was most convenient for the interviewee. The paper is structured as follows. Section 2 presents the background of this work. Section 3 introduces the research setup, and Section 4 lists the key findings. Section 5 discusses the key findings. Section 6 draws some final conclusions.

\section{Background and Motivation}


The Public Procurement Act \cite{directive-procurement} governs public acquisitions. However, it does not apply when a contracting authority, for example, a municipality, makes a procurement from a company it owns, called an in-house company, provided that the in-house company is formally separate for policy-making purposes, has a controlling interest by the municipality and conducts only a limited amount of business with external parties \cite{Finlex_2016}. Procurement Directive allows 20 percent of turnover to go outside the owners of the in-house company \cite{directive-procurement}. However, in Finnish law, the threshold for outselling is stricter. Public Procurement Act specifies that 5 percent and EUR 500,000 limits for outselling apply based on the in-house entity's turnover three years before the agreement \cite{Finlex_2016}. However, these limits don't apply when there's no market-based operation to execute the services. Whether the market-based operations exist is determined by the responses to a transparency declaration \cite{Finlex_2016}.

Procurement units that own the in-house company must have decisive authority in the in-house company \cite{Finlex_2016}. The Public Procurement Act defines joint-decisive authority as when all contracting entities have representatives in the in-house company's executive organs and collectively make strategic decisions, with the condition that the in-house company operates in the interests of the controlling contracting entities \cite{Finlex_2016}. In addition, the Public Procurement Act states that it does not apply when an in-house company is a procurement unit itself and procures goods or services from another procurement unit, which exercises controlling interest in the in-house company or another entity under the same controlling interest \cite{Finlex_2016}. This option is the so-called in-house sisters arrangement in Finland. The recent judgment of the EU Court of Justice (ECJ) in the Sambre \& Biesme case \cite{Sambre_biesme} would seem to contradict the article in the Finnish Public Procurement Act or at least guide how to interpret Section 15 of the Procurement Act. In this case, the need for real representation in the in-house company's board or decision-making bodies was emphasized, possibly contradictory to the Procurement Act. Ownership of shares alone did not guarantee decisive authority in the in-house sister arrangement, even if the other procurement unit had decisive authority in the in-house company. This shows that factors related to the in-house company's governance and joint-decisive authority can significantly impact assessing its in-house status.


Some other ECJ judgments depict how to evaluate adequate in-house position. In the Parking Brixen case \cite{Parken_brixen_2005}, the municipality lacked sufficient decisive authority in the in-house company, rendering the company not part of the municipal in-house. Similarly relating to the evaluation of the owner's sufficient decisive authority, the Carbotermo and Concorzia Alise case \cite{Carbotermo_consorzio_2006} considered how the strong dominant position of majority shareholder affects the legal position of other shareholders in an in-house company. The risk for conflict of interests is high, and it can influence the in-house company's legal position. If only one or a few shareholders have real decisive authority, the objectives of the other owners are not given space; their realization is uncertain and, therefore, it may create a situation where those with little or no decisive authority do not have a real in-house position in the company they own.


A typical situation in the Finnish in-house landscape is that the in-house position is seen as a habitual practice through ownership and through a somewhat fictitious demonstration of decisive authority. Within similar themes, in Econord case, the significance of structural and operational control in assessing in-house status was highlighted \cite{Econord_2012}. Mere formal ownership is insufficient to ensure in-house status \cite{Econord_2012}. This suggests that even small shareholders should have sufficient joint-decisive authority over the in-house company's operations, and in-house position cannot be presented merely on paper. 
For example, the largest Finnish in-house company, Kuntien Tiera, has 398 owners. As methods of decisive authority, Kuntien Tiera states that the owners steer Kuntien Tiera's activities in the general assembly and the board of directors, as well as the developing Kuntien Tiera's service offerings in six different steering groups \cite{Tiera_2023}.


Based on these legal cases, it is evident that the importance of real decisive authority and ownership in the in-house company is significant. In addition to ownership share, importance is also given to control, structure, decision-making, and genuine representation in the in-house company's operations. It is important to assess these factors as a whole when evaluating the legal status of an in-house company.


The in-house arrangement can be challenging to interpret for municipalities \cite{ghezzi2022role}. Despite clear guidelines provided by case law, there is a significant variation in their interpretation in practice \cite{ghezzi2022role}. The legal setup surrounding in-house procurement is a critical issue discussed in the literature. In Poland, where stricter in-house procurement criteria have been implemented, the debate is polarised between supporters and opponents \cite{hartung2018in}. Opponents seem to question whether in-house practice aligns with the goals set in legislation \cite{hartung2018in}. Similarly, Burgi and Koch \cite{burgi2012in} evaluate the Public Procurement Directive article 11 and suggest that lowering the criteria for in-house procurement could be a way to prevent legal mismatch and confusion in the field.

In practical applications, in-house procurement may benefit smaller municipalities by reducing the bureaucracy involved in contracting and contract implementation costs \cite{miemiec2013application}. However, it has been questioned whether the upcoming, now-current directives will create a procurement market that does not have to obey and is not controlled by procurement norms \cite{hausmann2013house}. The concerns are that the upcoming directives will exclude private service providers from the competition if the in-house exception is accepted \cite{hausmann2013house}. Similar concerns have been raised in Finland as well. The Confederation of Finnish Industries has raised concerns that the current in-house practice distorts the market and has taken steps to address these concerns through a request for measures to the practices from the Competition and Consumer Authority \cite{Ek_lausunto_2023}. Baciu suggests that public bodies should not be able to avoid transparent procedures and contract directly with other public bodies, except in rare and limited situations to preserve fair competition \cite{baciu2015horizontal}. The Confederation of Finnish Industries and the Finnish Competition and Consumer Authority also take the same view in their proposals \cite{Ek_lausunto_2023,KKv_esitys_2023}. 

The literature concludes the current procurement directive inhibits opening up the national procurement markets and fosters direct awarding in public contracts, even if the underlying purpose is the opposite. The challenges surrounding in-house procurement for public entities highlight the need for continued examination and clarification of legal frameworks and in-house procurement criteria.

\section{Research Approach}

The research will focus on 
municipalities and well-being services counties in Finland. The research questions for this study are:

\begin{itemize}
    \item When should a public organization procure from in-house and when to procure from the market?
    \item What are the experiences of municipalities and well-being services counties with in-house ICT procurement?
    \item How much real decisive authority public sector organizations municipalities hold in the in-house arrangement?
\end{itemize}
\smallskip
\noindent
\textbf{Data Collection}.
The primary data collection method was semi-structured interviews with sixteen key stakeholders from municipalities and well-being services counties. The interviews were conducted face-to-face or via video conferencing. The approach to design the interviews was constructivist \cite{charmaz2014}, and therefore, the interviews were recorded because the aim was to preserve the details such as participant's tempo and tone as precisely as possible. However, only the audio of all interviews was recorded, and otherwise, for observation purposes, notes taken during the interview were relied upon. According to Glaser, the notes were able to capture what is needed without losing the detail \cite{glaser2001grounded}. During the analysis phase of this study, it was found that the recordings were an excellent supplement for interpreting the interviewee's attitudes and assumptions of in-house procurement. Especially when discussing more difficult topics, such as the legal status of in-house companies or the role of the small owner, the recordings helped to understand the hesitation and uncertainty. Only one interviewee requested that the video link not be used. Transcribed interview data was loaded into the \textit{atlas.ti} software for coding.

All participants were professionals in their field, either in public procurement in general, ICT procurement and its management, or in the financial management of the organisation. All participants were involved in in-house procurement in one way or another (Table \ref{tab:interview_data}).


\begin{table}[tp!]
\begin{center}
\caption{Interview participants.}
\scriptsize
\begin{tabular}{@{}ccccc@{}}
\textbf{Organization} &
  \textbf{Abreviation} &
  \textbf{Position} &
  \textbf{Field} &
  \textbf{Duration} \\ 
Participant 1          & P1   & Chief Financial Officer   & Administration               & 107 \\
Participant 2          & 107  & Procurement Manager & Procurement / ICT               & 49 \\
Participant 3          & P3  & City Director         & Administration & 53 \\
Participant 4          & P4   & Head of Procurement Expert Group    & Procurement / ICT               & 49 \\
Participant 5          & P5   & Chief Digital Officer     & ICT       & 67 \\
Participant 6 & P6  & City Auditor     & Administration               & 57 \\
Participant 7 & P7 & Division Director   & ICT               & 51 \\
Participant 8 & P8 & Procurement Specialist  & Procurement / ICT               & 56 \\
Participant 9 &
  P9 &
 Procurement Manager &
 Procurement &
  56 \\
Participant 10 & P10 & Support Services Director   & ICT               & 75 \\
Participant 11 & P11  & Chief Information Officer (CIO)     & ICT               & 53 \\
Participant 12 & P12  & Chief Information Officer (CIO)   & ICT               & 61 \\
Participant 13 & P13 & Welfare County Director   & Administration               & 57 \\
Participant 14 & P14  & Municipal Director     & Administration               & 105 \\
Participant 15 & P15  & Administrative Director   & Administration              & 105 \\
Participant 16 & P16  & Chief Financial Officer   & Administration               & 105
\end{tabular}
\vspace{-6mm}
\end{center}
\label{tab:interview_data}
\end{table}

\smallskip
\noindent
\textbf{Analysis}.
The grounded theory (GT) approach suits topics lacking relevant research or where a new perspective is desired \cite{urquhart2023}. The practice of ICT in-house procurement is an unexplored area in Finland, except for the request for measures \cite{KKv_esitys_2023} and report \cite{KKv_selvitys_2021} by the Consumer and Competition Authority and surveys conducted by Confederation of Finnish industries \cite{Lith_2022}. Fresh European in-house procurement research is also extremely limited. 

The GT approach to research involves systematically coding and classifying data \cite{strauss1990basics}. GT stands apart from other qualitative research methods primarily in its approach to analysis, while data collection methods can vary. Typically, GT involves constructing theories based on interview data, with data collection continuing until saturation is reached \cite{urquhart2023}. Saturation means that no new information relevant to the developing theory is emerging \cite{dey2004grounded}.

In this research, the coding followed a constructive approach to the grounded theory \cite{charmaz2014}. The open coding stage included initial coding and sometimes codes that emerged from the participants' narratives, known as "in vivo" coding. This constituted the first analysis phase, establishing a data-driven initial sorting \cite{charmaz2014}. The initial codes facilitated comprehension of the interview material and the intended meanings conveyed by the interviewees. Subsequently, after each interview, a comprehensive review of the material and codes was conducted to verify that the codes consistently conveyed the same concept across all interviews. Charmaz underscores the significance of constant comparison within GT, a methodology involving the comparison of categorized data instances within the same category \cite{charmaz2014}. As outlined by Urquhart in 2023, this approach aims to assess the compatibility and efficacy of the identified categories \cite{urquhart2023}.   

As coding progressed in the study, focused coding advanced the analysis to a more theoretical direction with conceptualization, for example, recognizing where the initial codes lead the process: 
\begin{quote}
    "\textit{Feels disempowered in cooperation.}" -- "\textit{Signs of insufficient decisive authority.}"
\end{quote}
After focused coding, thoughts arose about the relationships between these codes. These relationships were marked utilizing the \textit{atlas.ti} memo and grouping function. At this point, the axial coding stage \cite{charmaz2014} and the selective coding stage were somewhat parallel processes \cite{urquhart2023} \cite{charmaz2014}. The phase of seeking common themes and grouping categories helped us understand the causation relationships. 

The significance of theoretical notes in understanding relationships was emphasized and aided in forming an overall picture. Coding, categorization, and grouping were flexible throughout the analysis, and changes occurred until the key categories were fully saturated and no new codes emerged. Ultimately, 996 quotations were selected from the material and categorized under 149 codes. It should be noted that around 700 additional quotations were coded related to clusters, such as themes concerning the organization of public entities in procurement, monitoring and measurement of procurement, ICT project management, public organization management, and system solution-related themes. 


\section{Results}

\subsection{Reasons for ICT In-House Procurement}

There are several characteristics that where in-house procurement can be justified. It allows sharing the risk and costs of producing certain widely used services, as well as due to different financial capacities of public sector organizations. Below, we present the key reasons for using ICT in-house procurement found in this study.


\noindent
\textbf{ICT in-house companies are widely utilized due to shortcomings in the existing market.} 
At times, only few (and sometimes no) bids are received for ICT procurement. Then, in-house companies are the sole providers capable of offering support services to the public sector organizations, such as systems for managing human resources and payroll. Municipalities and welfare service counties believe it would be a welcome addition if market players extended their services to the sector where ICT in-house companies currently operate. Available solutions and service production encounter challenges that are believed to be alleviated through increased competition within the sector, thereby providing alternative solutions to meet various needs.

In addition, interviews reveal that ICT in-house companies are extensively utilized for ICT hardware and equipment procurement, even though this type of procurement is typically considered straightforward. Some public organizations procure equipment through in-house channels, driven by the belief that the market cannot provide the necessary volumes. However, certain public organizations have come to realize that ICT equipment obtained through in-house procurement tends to be more expensive than market-based solutions. These organizations emphasize that entities should explore what markets can offer to ensure the most responsible use of public funds. 

\smallskip
\noindent
\textbf{ICT in-house procurement is faster than competitive bidding.}
Obtaining products and services from an ICT in-house company is a straightforward process. Local government sectors often have limited resources to engage in bidding, which typically occurs alongside employees' regular duties, often in collaboration with the procurement team or center. However, expertise must come from within the specific sector to oversee the bidding process.

ICT in-house procurement can enhance municipal operations by efficiently utilizing resources, time, and expertise required for daily operations. In comparison to competitive bidding, ICT in-house procurement is swift and convenient for municipalities, especially for fulfilling simple needs. Interviews also underscore that ICT in-house procurement is considered a reliable method, particularly in smaller organizations where the likelihood of legal disputes is reduced. Competitive bidding is seen as burdensome and error-prone, making ICT in-house procurement a suitable option, particularly when resource constraints are a factor.

Finally, ICT in-house procurement played a pivotal role in the recent establishment of well-being services in counties instead of municipalities, which had organized the services previously. The timeline was so strict that would have been impossible to tender market-based competitive bidding, as per procurement law. Furthermore, central procurement organizations lacked the capacity for proper competitive bidding while the establishment of well-being services in counties was under construction. Then, through ICT in-house companies, well-being services in counties were operationalized within a tight 1.5-year timeframe.

\smallskip
\noindent
\textbf{Resources and expertise within public organizations may often prove inadequate.} More than half of the interview participants believe that public organizations lack personnel who understand the ICT needs of the sectors well enough to support the creation of coherent system configurations. Additionally, these organizations often lack personnel who can simultaneously grasp the diverse requirements of competitive bidding in accordance with procurement laws. When a public organization lacks both ICT and procurement expertise, ICT in-house procurement becomes a viable option for acquiring products and services simply because everything seems to be readily available off-the-shelf.

\smallskip
\noindent
\textbf{The desire is to centralize collaboration in one location and obtain shared and standardized ICT systems through in-house procurement.}
Local governments and well-being services counties believe that certain needs within public organizations are quite similar, particularly those related to support services. Municipalities seek to harness the benefits of collaboration and shared systems to achieve cost-efficiency and agility in such cases. Furthermore, system compatibility among municipalities is perceived as facilitating rapid service delivery and error correction. The ICT in-house practice may not always meet this need, leading some municipalities to purchase the same system offered by ICT in-house directly from the system provider in an attempt to resolve issues directly with the supplier.

\smallskip
\noindent
\textbf{ICT in-house procurement is needed to enhance collaboration.}
ICT in-house companies have emerged because clear, distinguishable functions within public organizations are identified for collaborative production with other entities that share similar needs. An example of such a function could be payroll processing. The goal is to enhance the efficiency of public organizations by centralizing and sharing production costs with other stakeholders while freeing up internal resources. Additionally, centralization aims to harness expertise-related benefits, allowing for the incorporation of necessary expertise from external sources, where such expertise is perceived to be concentrated within that specific function. The ICT in-house practice also aims to ensure the security of critical system operations and their continuous functionality.

\subsection{Key Problems Related to ICT In-House Companies}

Despite the benefits, some problems arise in the context of ICT in-house companies. Table \ref{tab:issues} provides an overview of key issues related to ICT in-house companies. In summary, insufficient decisive authority, the position of minority shareholders, rapid expansion of ICT in-house companies, damaged reputation, costly solutions, deficiencies in contract practices, and issues related to ownership shares emerge as central problems based on the study. This section discusses the challenges within ICT in-house companies and their potential sources.

\begin{table}[tp!]
\begin{center}

\scriptsize
\caption{Key issues in in-house procurement}

\begin{tabular}{|l|l|}
\hline
\textbf{\begin{tabular}[c]{@{}l@{}}Key issues in in-house procurement in this research.\end{tabular}} &
  \textbf{\begin{tabular}[c]{@{}l@{}}Percentage of how \\many  interviewees \\ hold the opinion.\end{tabular}} \\ \hline
\begin{tabular}[c]{@{}l@{}}No real decisive authority in in-house company\end{tabular}                        & 81 \% \\ \hline
Small shareholder: Small buyer                                                                                   & 75 \% \\ \hline
\begin{tabular}[c]{@{}l@{}}In-house companies are currently too large\end{tabular}                             & 69 \% \\ \hline
Poor   reputation                                                                                                & 63 \% \\ \hline
Expensive solutions                                                                                            & 63 \% \\ \hline
In-house ownership, shareholder position                                                                                     & 63 \% \\ \hline
\begin{tabular}[c]{@{}l@{}}Insufficient expertise in the system  development and/or \\ procurement\end{tabular}   & 56 \% \\ \hline
\begin{tabular}[c]{@{}l@{}}Contracts with in-house companies are not binding or \\ they do not exist\end{tabular} & 44 \% \\ \hline
Exiting from in-house is demanding                                                                               & 44 \% \\ \hline
In-house: not functioning as it should                                                                           & 44 \% \\ \hline
Chain of command doesn't work                                                                                  & 38 \% \\ \hline
\begin{tabular}[c]{@{}l@{}}Service and system development are slow\end{tabular}                               & 31 \% \\ \hline
Poor quality of relationships                                                                                    & 31 \% \\ \hline
Trust has been eroded                                                                                            & 25 \% \\ \hline
\begin{tabular}[c]{@{}l@{}}Service does not meet the agreed  terms\end{tabular}                                & 25 \% \\ \hline
\end{tabular}

\label{tab:issues}
\end{center}
\end{table}

\begin{figure}[!t]
    \centering
    \includegraphics[width=0.75\textwidth]{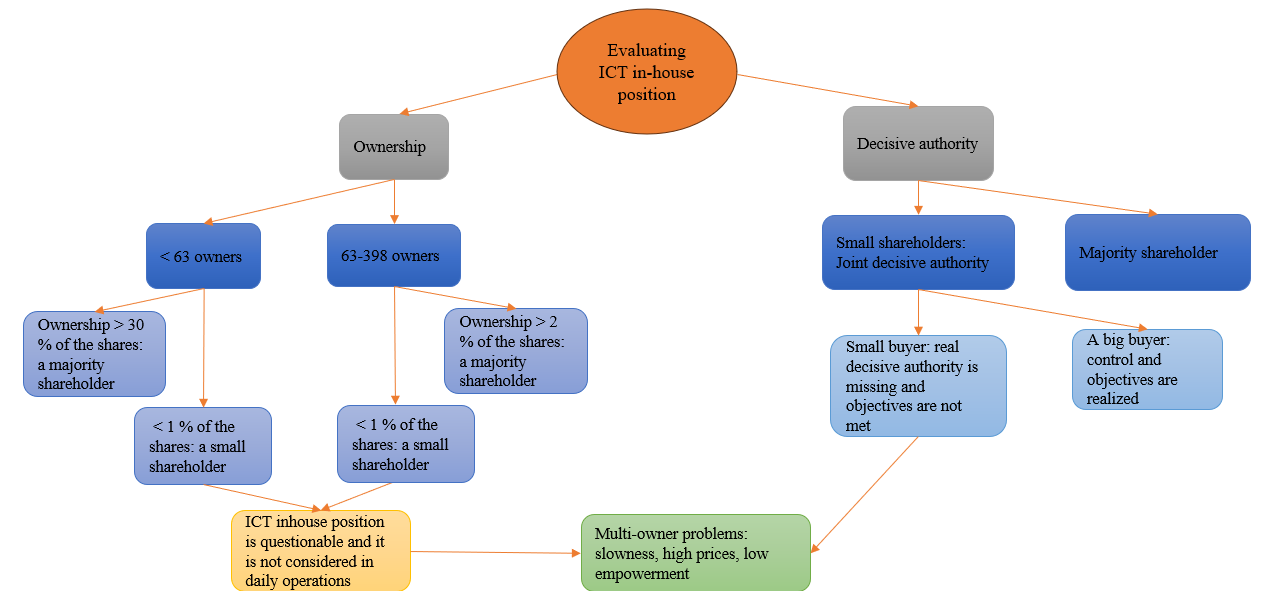}
    \caption{Evaluation of the in-house position in studied organisations.}
    \label{fig:inhouse_position}
\end{figure}

\smallskip
\noindent
\textbf{Challenges related to insufficient decision-making authority and the legal position of small shareholders.}
In municipalities and well-being services counties, there is a comprehensive understanding of how an in-house position could be achieved through procurement law. Ownership in the in-house company and decisive authority are central for the evaluation, as shown in Figure \ref{fig:inhouse_position}. All organizations in this research are small shareholders in the central in-house companies which we took for reference. Wide consensus exists about marginal ownership, seen as an established practice, and interviewees believe there is hardly room to interpret the matter differently. 

\begin{figure}[!t]
    \centering
    \includegraphics[width=0.75\textwidth]{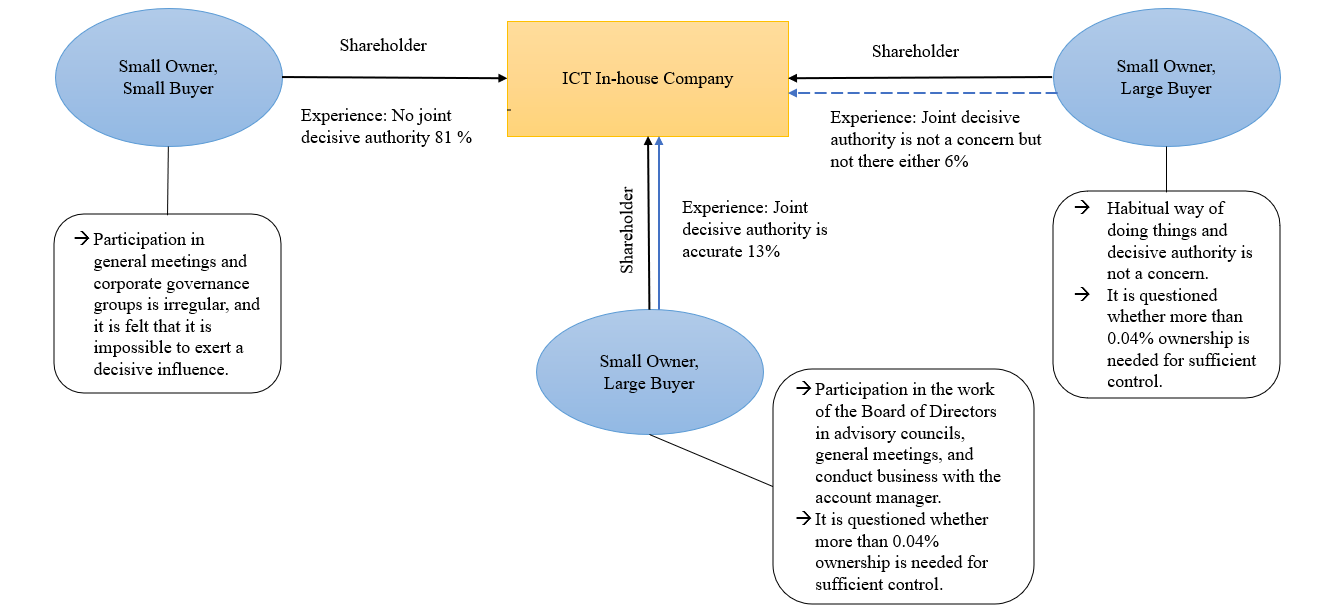}
    \caption{Recognized differences between minority shareholders' views about decisive authority and ownership.}
    \label{fig:owners}
\end{figure}

The problem arises from the unclear interpretation of sufficient decisive authority, which is also evident in interviews through varying interpretations. Within the interviews, three interpretations existed, as presented in Figure \ref{fig:owners}. Joint-decisive authority divides opinions. Most interviewees depict that  mechanisms work with even a small ownership stake, or nominal authority and a small ownership stake are deemed sufficient for the in-house position. The difference arises when considering the purchase sizes mentioned by interviewees. Large buyers feel that authority works and collaboration with ICT in-house companies is immediate. Problems are reacted swiftly, and organizational goals are achieved through in-house ICT collaboration. Some large buyers actively participate in decision-making bodies. One large buyer expressed thoughts about ownership not guaranteeing sufficient decisive authority:
\begin{quote}
   "\textit{To me, these shares and decisive authorities and such; the idea that ownership gives you certain position, I might not fully buy it. And then I think, are these matters as extensive as they have been portrayed in public.}" (P3)
\end{quote}

Some large buyers do not directly engage in the decision-making of ICT in-house companies, but they trust that shared authority is sufficient for evaluating the in-house position:
\begin{quote}
   "\textit{Well, there's a well-established legal practice in Finland that you don't need to think about it; if you have an in-house service provider and you've delved into it a bit, then you don't need a separate evaluation. Well-established legal practice means that there's such an in-house service provider where the owners exercise decisive authority together. The legislation is actually quite clear. It doesn't require any extraordinary evaluation. Of course, if the Competition and Consumer Authority ask, then we hire a lawyer who writes 10 pages about how it (joint-decisive authority) is done, but the matter is just this simple.}" (P1) 
\end{quote}

All small shareholders with significant purchases consider ICT in-house operations to align with their goals and find their authority in in-house companies effective. This is why the situation becomes problematic when we consider the experiences of small owners with small purchases. The views of large and small buyers are conflicting, as small buyers perceive there to be no real decisive authority in the ICT in-house companies:
\begin{quote}
   "\textit{Almost non-existent (decisive authority mechanisms). We own 0.01 of shares there, and then we're supposed to have decisive authority. If this counts as an in-house company as per procurement law, I've also thought a lot about how this can be.}" (H4)
\end{quote}
Again, in-house position is evaluated based on ownership and decisive authority, yet the small buyer's experience differs significantly from that of larger buyers. Consistently, small buyers question whether they possess a sufficient number of shares to attain proper decision-making authority within the in-house company, here we see how these two factors are assessed as equivalent criteria in determining the position of in-house companies, which differs from the reports of large buyers.

\begin{quote}
   "\textit{Well, the influence there is really small, that they are owner-managed companies, but each owner has such a small share that we don't know who actually controls it.}" (P5)  
\end{quote}

In addition, small buyers have refrained from participating in situations where joint decision-making authority could be demonstrated because it has been deemed futile:  
\begin{quote}
   "\textit{None of us have actually attended the general meetings anymore. Formally speaking, there are these owner meetings where strategic matters are discussed, where all over 100 shareholders use their weighty vote, and there's also a formal board member representing minority shareholders. I don't really feel that we have concrete influence over it (in-house company).}" (P6) 
\end{quote}

In summary, the majority of small shareholders with modest purchases believe that they lack significant authority over ICT in-house companies. Moreover, all study participants view ICT in-house companies as part of the market since the control mechanism does not function as intended for their own units. If the same objectives were applied to ICT in-house companies as for their own units, they could be considered an integral part of their own production.

\smallskip
\noindent
\textbf{Fast Expansion of the ICT In-House Companies.}
The interview responses suggest a significant increase in the number of owners of ICT in-house companies in recent years, largely due to mergers of smaller regional entities into larger national ones. This growth, particularly in the context of the central ICT in-house companies examined in the study, has been substantial, especially in terms of the number of minority shareholders.
The interviews also shed light on the challenges faced by minority shareholders, particularly those with smaller purchases, compared to majority shareholders. Notably, municipalities have observed that larger cities with greater ownership and purchasing power tend to receive priority in terms of the systems offered and their quality. This bias towards major owners often results in the goals of minority shareholders with limited influence within the in-house company not being met.
As a consequence, the existence of multiple owners poses considerable challenges in achieving common objectives. In the central ICT in-house companies studied, as well as those discussed in the interviews, the ownership structure varies widely, ranging from 47 to 398 owners. It is noteworthy that all participating organizations hold a minority ownership position in these ICT in-house companies, with ownership stakes spanning from 0.00 to 1.00 percent of the shares.

\smallskip
\noindent
\textbf{Significant Variations in ICT In-House Companies' and Owner's Contract Practices.}
The study highlights significant variations in contract practices between ICT in-house companies and their owners. During the establishment of well-being services counties, some municipalities lacked contracts with ICT in-house companies, posing challenges when attempting to transfer contracts to the municipalities. Respondents also mention that the most significant problems with ICT in-house companies occur when contracts are entirely absent. Addressing errors becomes nearly impossible when the party supplying the system or service is not obligated to take any action.
In addition, uncertainties exist in contract clauses related to service levels, with a lack of specific obligations outlined for both the owner and the ICT in-house company. While most contracts state that problem situations should be resolved through collaboration, detailed service-level descriptions with obligations typical of the private sector seem to be entirely absent. Some ICT in-house companies prefer a standardized platform for all owner contracts that all owners can access, while others draft contracts only upon request.

\section{Discussion}

\textbf{Root Causes for Problems.}
Insufficient control by owners and the rapid expansion of ICT in-house companies are strongly interrelated. According to the study, there is an imbalance in the position of small shareholders, leading to problems associated with multi-ownership, such as the fact that small shareholders may not necessarily pursue common objectives. Small shareholders also hold very small ownership stakes, which raises the question of whether achieving dominant control in an ICT in-house company is structurally possible. If the interpretation is strict, the subsidiary status of ICT in-house companies might be problematic and contrary to the objectives of procurement law Section 15 \cite{Finlex_2016}.

Contractual practices vary a lot among in-house ICT companies and owners. Some ICT in-house companies have transparent contractual practices, while others have significant gaps in their contractual practices, leading to slow development of services and systems, difficulty in reacting to errors, and contracts lacking clear responsibilities for the in-house companies. ICT in-house companies dominate their market, and direct public competition rarely attracts many bids. The study indicates that 63 percent of the respondents consider solutions through in-house ICT companies expensive. However, municipalities and well-being services counties might not have any alternative but to continue with ICT in-house services, as migration costs would be too high. The lack of competition often results in price increases and decreased quality. Smaller owners are also forced to implement system updates and changes, which is relatively more expensive for them than for larger buyers. Table \ref{tab:problem_consequence} presents the recognized interrelationships. 

\begin{table}[tp!]
\begin{center}

\scriptsize
\caption{Antecedents, Field Experiences and Consequences.}
\begin{tabular}{|l|l|l|}
\hline
Antecedents A1-A4 &
  Experiences   E1-E8 &
  Consequences   C1-C5 \\ \hline
\multirow{3}{*}{\begin{tabular}[c]{@{}l@{}}A1. Fast expansion of \\ ICT in-house companies\end{tabular}} &
  \begin{tabular}[c]{@{}l@{}}E1. Lack of decision-making \\ power by the owners\end{tabular} &
  \multirow{3}{*}{\begin{tabular}[c]{@{}l@{}}C1.   Common objectives \\ are not met \end{tabular}} \\ \cline{2-2}
 &
  \begin{tabular}[c]{@{}l@{}}E2. Small owners and small \\ buyers have a weak position\end{tabular} &
   \\ \cline{2-2}
 &
  \begin{tabular}[c]{@{}l@{}}E3. In-house status \\ sometimes questionable\end{tabular} &
   \\ \hline
\multirow{3}{*}{\begin{tabular}[c]{@{}l@{}}A2. Shortcomings in \\ contractual practice.\end{tabular}} &
  \begin{tabular}[c]{@{}l@{}}E4.   Service and system \\ development is slow, reacting \\ to issues and errors is   slow.\end{tabular} &
  \multirow{3}{*}{\begin{tabular}[c]{@{}l@{}}C2 Current practice \\ does not hold \\ ICT in-house companies \\ liable for errors.\end{tabular}} \\ \cline{2-2}
 &
  \begin{tabular}[c]{@{}l@{}}E5. Vendor lock-in with   \\ in-house company and supplier.\end{tabular} &
   \\ \cline{2-2}
 &
  \begin{tabular}[c]{@{}l@{}}E6. Changes are almost   \\ impossible.\end{tabular} &
   \\ \hline
\begin{tabular}[c]{@{}l@{}}A3. ICT in-house \\ companies dominate \\ their market sector\end{tabular} &
  E7. Expensive solutions. &
  \begin{tabular}[c]{@{}l@{}}C3. Operations are \\ interrupted or significantly \\ impeded.\end{tabular} \\ \hline
\multirow{2}{*}{\begin{tabular}[c]{@{}l@{}}A4. The market is not \\ working / No competition \end{tabular}} &
  \multirow{2}{*}{\begin{tabular}[c]{@{}l@{}}E8. Vendor lock-in with \\ in-house company \\ and supplier. \end{tabular}} &
  C4. High costs \\ \cline{3-3} 
 &
   &
  \begin{tabular}[c]{@{}l@{}}C5. Updates and changes \\ are mandatory. \end{tabular} \\ \hline
\end{tabular}
\label{tab:problem_consequence}
\end{center}
\end{table}

\smallskip
\noindent
\textbf{Identified Preconditions for Success.}
When functioning properly, ICT in-house companies could bring efficiency, free up resources, and provide the necessary expertise to their owner organizations which is similar that Miemec stated \cite{miemiec2013application}. A prerequisite for this is that ICT in-house companies should be manageable, ensuring the necessary structural and operational control as mandated by the law, enabling effective control of their operations. This implies that in-house ICT companies should have fewer owners yet enough to achieve economies of scale. The current Finnish government has recommended that ownership shares in in-house companies should comprise a minimum of 10 percent. This proposal elicits apprehension regarding its possible detrimental impact on the well-established in-house model in Finland. More precisely, it has the potential to disrupt the current in-house structure, possibly encouraging the emergence of smaller, fragmented entities with duplicated responsibilities and management functions. Importantly, this may not necessarily foster the standardization of ICT systems and services.

One interessting option has not been studied. In the Sambre \& Biesme case, an in-house entity had different groups of owners with different decisive authority \cite{Sambre_biesme}. In the Finnish Limited Liability Companies Act, option to allocate decisive authority differently than \textit{one share -- one vote} principle is available as well \cite{Finlex_2006}. In this research we recognized different characteristics for different buyers and how join-decisive authority divides upon them. The shares in in-house companies are now allocated either according to the population base served by the owner organisation or in the cases of well-being services counties, we did not find the justification. The purchaser groups, whether the buyer is small or large, could help to even out or create new mechanisms how the decision-making should happen in the in-house company. This suggestion, however, needs more research to see whether it could be viable option in practice.  

\smallskip
\noindent
\textbf{Recommendations.}
This study identifies mechanisms that could enhance current in-house practices and improve public sector organisations' and market actors' influence over the operations of ICT in-house companies. In the literature \cite{burgi2012in}, it has been suggested that criteria for in-house procurement should be relaxed to avoid legal incompatibility and confusion. However, this study proposes a different approach since there is a lack of oversight and competition, resulting in significant national economic problems. The study reveals that the majority of respondents perceive control over ICT in-house companies to be weak, leading to slow development of services and systems, high costs, and challenges in correcting errors. The results suggest that, in certain situations, problems related to delivery can be avoided. In situations where ICT in-house companies are under the immediate control of their owners and control is closely aligned with the owners' goals, ICT in-house companies can serve as a resource to free up procurement competition. Close ownership relationships require sufficient ownership and less than fifty owners, enabling genuine structural and operation control. As a result, the procurement law needs clarification on what constitutes sufficient ownership in an in-house company. Contrary to \cite{burgi2012in}, our results indicate that clear control mechanisms, strong control, and evidence of in-house status from procurement law could help reduce legal incompatibility and confusion in in-house procurement. 

\smallskip
\noindent
\textbf{Threats to validity.}
While GT is considered data-driven, it is impossible to completely eliminate the influence of the researcher's prior experiences and theoretical frameworks. These factors inevitably shape the analysis. Moreover, for research to be meaningful, it should connect to previous studies and ongoing scientific discussions. Instead of strictly adhering to inductive or deductive reasoning, GT incorporates an intermediate approach known as abduction or theory-binding. This acknowledges the role of the researcher's thinking while recognizing the importance of existing theoretical tools and context.

\section{Conclusions}
In conclusion, in-house procurement remains a controversial issue in public procurement. While some argue that it provides flexibility and cost savings for public authorities, others express concern about potential abuses of the exemption and the impact on fair competition. As reflected, the legal framework surrounding in-house procurement is complex and subject to ongoing debate. 

This paper identified a range of key reasons for ICT in-house procurement and why it is important for its owners. Key problems were highlighted, and recommendations were formulated based on literature and research on practically improving operations. The research revealed valuable insights into the complex relationships between Finnish municipalities and their in-house companies. The study also touched upon the legal framework related to ICT in-house procurement, a pivotal issue in scholarly literature, emphasizing the ongoing need to review legal frameworks and in-house procurement criteria to address challenges posed to municipalities by in-house procurement.

%

%
%
%
\bibliographystyle{splncs04}
\bibliography{lahteet}

\end{document}